\documentclass[11pt,a4paper]{article}
\usepackage{jheppub_kim}

\usepackage{pdflscape}
\usepackage{amsmath}
\usepackage{amssymb}
\usepackage{dcolumn}
\usepackage{bm}
\usepackage{color}
\usepackage{epsfig}
\usepackage{amsfonts}
\usepackage{graphicx}
\usepackage{subfigure}
\usepackage{dcolumn}

\newcommand{\be}{\begin{equation}}
\newcommand{\ee}{\end{equation}}
\newcommand{\bea}{\begin{eqnarray}}
\newcommand{\eea}{\end{eqnarray}}

\def\be{\begin{equation}}
\def\ee{\end{equation}}
\def\bea{\begin{eqnarray}}
\def\eea{\end{eqnarray}}

\begin{document}

\title{Extended Chaplygin Gas in Horava-Lifshitz Gravity}

\author[a]{B. Pourhassan}

\affiliation[a]{School of Physics, Damghan University, Damghan, Iran}

\emailAdd{b.pourhassan@du.ac.ir}

\abstract{In this paper, we investigate cosmological models of the extended Chaplygin gas in
a universe governed by Horava-Lifshitz gravity. The equation of state for an extended Chaplygin gas is a $(n+2)$-variable equation
determined by $A_{n}$, $\alpha$, and $B$. In this work, we are interested to the case of second order $(n=2)$ equation of state which recovers quadratic barotropic equation of state. In that case there are four free parameters. We solve conservation equation approximately and obtain energy density in terms of scale factor with mentioned free parameters. Under some assumptions we relate free parameters to each other to have only one free independent parameter $(A_{2})$. It help us to obtain explicit expression for energy density in terms of scale factor. The allowed values of the second order extended Chaplygin gas parameter is fixed using the recent astrophysical and cosmological observational data. Thermodynamics of the model investigated based on the first and second law of thermodynamics.}


\maketitle

\section{Introduction}
Cosmological observations verify the accelerated expansion of the universe, including the deceleration to the accelerated phase transition \cite{Riess:1998cb,Riess:2004a,Perlmutter:1998np}. Such accelerated expansion can be described by dark energy models.
One of the primordial candidates of dark energy is the cosmological constant, which is not dynamical model, so there are some alternative models such as quintessence model \cite{Ratra:1987rm,Wetterich:1987fm,Liddle:1998xm,Guo:2006ab,Khurshudyan:2014a,Dutta:2009yb}, phantom model \cite{Caldwell:1999ew,Caldwell:2003vq,Nojiri:2003vn,Onemli:2004mb,
Saridakis:2008fy,Saridakis:2009pj,Gupta:2009kk}, or quintom model \cite{Guo:2004fq,Zhao:2006mp,Cai:2009zp}. Holographic model of dark energy is another interesting description of the dark energy \cite{Li:2010a,Sadeghi:2014a,Setare:2007azp,Saridakis:2008azp}.\\
In order to construct dynamical version of cosmological constant model, one can introduce interaction term between dark matter and dark energy \cite{P45,P54,P49,P58,P48,P74,Boehmer:2008av,P61,Chen:2008ft,P50}.\\
There are also other interesting models to describe the dark energy such as Chaplygin gas \cite{kamenshchik,Bento:2002ps}, which emerged initially in cosmology from string theory point of view \cite{john1,john2}, which are based on Chaplygin gas (CG) equation of state and developed to the generalized Chaplygin gas (GCG) \cite{P83}.
It is also possible to enter the presence of viscosity in GCG \cite{P86,P87,P88,P91,P89}.
Then, GCG was extended to the modified Chaplygin gas (MCG) \cite{P92}. Recently, viscous MCG is also suggested and studied \cite{P93,P94}. A further extension of CG model is called modified cosmic Chaplygin gas (MCCG) which was proposed recently \cite{P96,P97}.\\
The MCG equation of state (EoS) has two parts, the first term gives an ordinary fluid obeying a linear barotropic EoS, and the second term relates pressure to some power of the the inverse of energy density. So, one essentially dealing with a two fluid model. However, it is possible to consider barotropic fluid with quadratic EoS or even with higher orders EoS \cite{P100-1,P100-2,P100-3}. Therefore, it is interesting to extend MCG EoS which recovers at least barotropic fluid with quadratic EoS, and is called extended Chaplygin gas (ECG) \cite{P102,P103,P1002,P1003}.\\
On the other hand, Horava-Lifshitz (HL) gravity appears
to be an attractive model to achieve a complete quantum gravitational theory \cite{H-L1}. As we know, there are several open issues in HL gravity including the classical and quantum instability of the scalar modes which can be solved under some assumptions like consideration of projectable HL gravity and principle of detailed balance. In that case, HL gravity in the presence of a scalar field studied by the Ref. \cite{GC} where the effect of detailed balance conditions investigated. Moreover it is mentioned that HL theory of gravity is not exactly consistent with general theory of relativity at low energy \cite{KM}.\\
HL gravity has some application in the black hole properties \cite{Bh1,Bh2,Bh3,Bh4,Bh5,Bh6}, the thermodynamic
properties \cite{TBH1,TBH2,TBH3,TBH4,TBH5}, the dark
energy phenomenology \cite{DE1,DE2,DE3}, etc. Additionally,
application of HL gravity as a cosmological framework gives rise
to HL cosmology \cite{HLC1,HLC2}.\\
In that case, HL cosmology with GCG has been studied by the Ref. \cite{HLGCG1}. Also, MCG in HL gravity and observational constraints has been studied by the Refs. \cite{HLMCG1,HLMCG2} with possibility of extension to the case of varying $G$ and $\Lambda$ \cite{HLMCG3}. Now, we would like to investigate ECG in HL gravity. The equation of state for a ECG is a $n+2$-variable equation
determined by $A_{n}$, $\alpha$ and $B$. We assume second order EoS which recovers quadratic barotropic EoS, and by using special assumptions, reduce number of free parameters to one. Therefore, we have only an independent parameter. The allowed values of mentioned EoS parameter are fixed by using the recent
astrophysical and cosmological observational data such as $H(z)$ analysis.\\
On the other hand, the temperature behavior and the thermodynamic stability of the generalized Chaplygin gas has been studied by the Ref. \cite{PLB636(2006)86}, and it is found that the generalized Chaplygin gas cools down through the expansion
without facing any critical point or phase transition.  Also, thermodynamics of the generalized Chaplygin gas has been investigated by introducing the integrability condition, and thermodynamic quantities have been derived as functions of either
volume and temperature \cite{0812.0618}. Validity of the generalized second law of gravitational thermodynamics in a non-flat Friedmann-Robertson-Walker (FRW) universe and an expanding G\"{o}del-type universe containing the generalized Chaplygin gas confirmed by the Refs. \cite{P26 1103.4842} and \cite{IJTP52(2013)4583} respectively. In the extension of the Ref. \cite{PLB636(2006)86}, the similar work performed for the case of the modified Chaplygin gas \cite{PLB646(2007)215} and the same result obtained. More discussion on thermodynamical behavior of the modified Chaplygin gas found in the Ref. \cite{P58 1006.1461}. Also, Ref. \cite{P59 1012.5532} developed the Ref. \cite{0812.0618} to the case of the modified Chaplygin gas. Validity of the generalized second
law of thermodynamics in the presence of the modified Chaplygin gas investigated by the Ref. \cite{P60 1102.1632} and observed that the generalized second law of thermodynamics always satisfied for the modified Chaplygin gas model. The generalized second law of thermodynamics in the brane-world scenario including the modified Chaplygin gas verified on the apparent horizon in late time by the Ref. \cite{ASS341(2012)689}.\\
Already, the thermodynamics in HL cosmology have been investigated and validity of the generalized second law of thermodynamics verified \cite{TBH5}. So, there is no thermodynamical study of the ECG in HL cosmology, which is also another subject of this paper.\\
The paper is organized as follows. In the next section we give a brief review of HL cosmology, then in section 3 we introduce the ECG. In section 4 we use observational data to constrain the free parameters of the model. In section 5 we investigate some cosmological parameters and in section 6 we study thermodynamical aspects of the model and finally in section 7 we give conclusions and suggestions for future works.

\section{Horava-Lifshitz cosmology}
HL gravity described by the following metric \cite{HLGCG1},
\begin{equation}\label{HL1}
ds^{2}=-N^{2}dt^{2}+g_{ij}(dx^{i}+N^{i}dt)(dx^{j}+N^{j}dt),
\end{equation}
where $N$ and $N^{i}$ are the lapse
and shift functions  which are used in general relativity in order to
split the space-time dimensions. Using the projectable version of HL gravity \cite{Project1} with the detailed balanced
principle \cite{H-L2} one can write the gravity action of HL theory as follow,
\begin{equation}\label{HL2}
S=\int{dt dx^{3}\sqrt{g}N[\tilde{\mathcal{L}}_{0}+\mathcal{L}_{0}+\mathcal{L}_{1}]},
\end{equation}
where,
\begin{eqnarray}\label{HL3}
\tilde{\mathcal{L}}_{0}&=&\frac{2}{\kappa^{2}}(K_{ij}K^{ij}-\lambda K^{2}),\nonumber\\
\mathcal{L}_{0}&=& -\frac{\kappa^{2}}{2\omega^{4}}C_{ij}C^{ij}-\frac{\kappa^{2}\mu}{2\omega^{2}}\frac{\epsilon^{ijk}}{\sqrt{g}}R_{il}\nabla_{j}R_{k}^{l}
-\frac{\kappa^{2}\mu^{2}}{8}R_{ij}R^{ij},\nonumber\\
\mathcal{L}_{1}&=&\frac{\kappa^{2}\mu^{2}}{8(1-3\lambda)}\left(\frac{1-4\lambda}{4}R^{2}+\Lambda R-3\Lambda^{2}\right),
\end{eqnarray}
where $\kappa$, $\lambda$, $\mu$ and $\omega$ are constant
parameters, $\Lambda$ is a positive constant, which as usual is related to
the cosmological constant in the IR limit, $R_{ij}$ and $R$ are Ricci tensor and Ricci
scalar respectively. Also, the Cotton tensor
is defined as follow,
\begin{equation}\label{HL4}
C^{ij}=\frac{\epsilon^{ijk}}{\sqrt{g}}\nabla_{k}(R_{i}^{j}-\frac{1}{4}R\delta_{i}^{j}),
\end{equation}
also the extrinsic curvature is defined as,
\begin{equation}\label{HL5}
K_{ij}=\frac{1}{2N}(\dot{g}_{ij}-\nabla_{i}N_{j}-\nabla_{j}N_{i}).
\end{equation}
It is usual to use FRW metric with $N=1$ and $N^{i}=0$ to obtain the following Friedmann equations,
\begin{equation}\label{HL6}
H^{2}=\frac{\kappa^{2}}{6(3\lambda-1)}\rho-\frac{\kappa^{4}\mu^{2}}{8(3\lambda-1)^{2}}\left(\frac{\Lambda k}{a^{2}}-\frac{k^{2}}{2a^{4}}-\frac{1}{2}\Lambda^{2}\right),
\end{equation}
and,
\begin{equation}\label{HL7}
\dot{H}+\frac{3}{2}H^{2}=-\frac{\kappa^{2}}{4(3\lambda-1)}p-\frac{\kappa^{4}\mu^{2}}{16(3\lambda-1)^{2}}\left(\frac{\Lambda k}{a^{2}}+\frac{k^{2}}{2a^{4}}-\frac{3}{2}\Lambda^{2}\right),
\end{equation}
where $H=\dot{a}/a$ and $a$ are Hubble parameter and scale factor respectively, $k$ is curvature constant corresponding to open ($k<0$), flat ($k=0$), and closed ($k>0$) universe. Also, $p$ and $\rho$ are corresponding to total pressure and energy density which contain radiation, dark matter and dark energy.\\
Using the usual notifications,
\begin{equation}\label{HL8}
G_{cosmo}=\frac{\kappa^{2}}{16\pi(3\lambda-1)},
\end{equation}
\begin{equation}\label{HL9}
G_{grav}=\frac{\kappa^{2}}{32\pi},
\end{equation}
and,
\begin{equation}\label{HL10}
\frac{\kappa^{4}\mu^{2}\Lambda}{8(3\lambda-1)^{2}}=1,
\end{equation}
the conservation equations are,
\begin{equation}\label{HL11}
\dot{\rho}_{r}+3H(p_{r}+\rho_{r})=0,
\end{equation}
\begin{equation}\label{HL11-1}
\dot{\rho}_{b}+3H\rho_{b}=0,
\end{equation}
and,
\begin{equation}\label{HL12}
\dot{\rho}_{c}+3H(p_{c}+\rho_{c})=0,
\end{equation}
where $p_{r}$ and $\rho_{r}$ are pressure and energy density of radiation, $\rho_{b}$ is energy density of baryonic matter which is pressureless, and $p_{c}$ and $\rho_{c}$ are pressure and energy density of the extended Chaplygin gas which is the unification of the dark matter and dark energy and discussed in the next section.
\section{Extended Chaplygin gas}
The extended Chaplygin gas EoS given by \cite{P102,P103},
\begin{equation}\label{ECG1}
p_{c}=\sum_{n=1}^{\infty}A_{n}\rho_{c}^{n}-\frac{B}{\rho_{c}^{\alpha}},
\end{equation}
where $A_n$, $B$, $\alpha$ and $n$ are constants, so we have generally $n+2$ free parameters. Note that in the case $n=1$ the above
expressions recovers the standard MCG. In this paper, we are interested to the case of second order EoS ($n=2$) which recovers quadratic barotropic EoS. In that case the EoS given by (\ref{ECG1}) reduced to the following expression,
\begin{equation}\label{ECG2}
p_{c}=A_{1}\rho_{c}+A_{2}\rho_{c}^{2}-\frac{B}{\rho_{c}^{\alpha}}.
\end{equation}
Assuming $A_{2}=0$ gives MCG \cite{HLMCG1}, while  $A_{2}=A_{1}=0$ gives GCG \cite{HLGCG1} where $A_{1}$, $B$, and $\alpha$ are positive constants with $0<\alpha\leq1$. Therefore, we have four free parameters in our model.\\
Using the conservation equation (\ref{HL12}) and ECG equation of state (\ref{ECG2}) we can obtain,
\begin{equation}\label{R2}
\rho_{c}=\left[\frac{B}{1+A_{1}}+\frac{c}{a^{3(1+A_{1})(1+\alpha)}}e^{-(1+\alpha)(1+A_{1})f(\rho_{c})}\right]^{\frac{1}{1+\alpha}}
\end{equation}
where $c=(1+A_{1})^{-1}$ and
\begin{eqnarray}\label{R3}
f(\rho_{c})&=&\frac{A_{2}\rho_{c}}{(1+A_{1})^{2}}-\frac{BA_{2}\rho_{c}}{(1+\alpha)(1+A_{1})^{2}((1+A_{1})\rho_{c}^{1+\alpha}-B)}\nonumber\\
&+&A_{2}\int{\frac{B(2+\alpha)}{(1+\alpha)(1+A_{1})^{2}((1+A_{1})\rho_{c}^{1+\alpha}-B)}d\rho_{c}}.
\end{eqnarray}
In the case of $A_{2}=0$, then $f(\rho_{c})=0$, so we recover the result which is obtained by MCG. In order to isolate $\rho_c$ completely and find correct relation of $\rho_c$ in terms of $a$, we consider special case of $\alpha\ll1$, in that case integration of last term in (\ref{R3}) is solved analytically and we find,
\begin{equation}\label{R4}
\int{\frac{B(2+\alpha)}{(1+\alpha)(1+A_{1})^{2}((1+A_{1})\rho_{c}^{1+\alpha}-B)}d\rho_{c}}|_{\alpha\rightarrow0}=\frac{2B}{(1+A_{1})^{3}}\ln{((1+A_{1})\rho_{c}-B)}.
\end{equation}
Asymptotic behavior (early universe where $\rho_{c}$ is very big) together $a\ll1$ allow us to choose,
\begin{equation}\label{R5}
ce^{-(1+\alpha)(1+A_{1})f(\rho_{c})}\approx1+C(A_{2})\rho_{c}^{-(1+\alpha)},
\end{equation}
where $C(A_{2})$ is a constant depend on $A_{2}$ which is zero at $A_{2}=0$. Under these assumptions one can obtain,
\begin{equation}\label{R6}
\rho_{c}^{1+\alpha}=\frac{1}{2}\left(\frac{B}{1+A_{1}}+\frac{1}{a^{3(1+A_{1})(1+\alpha)}}\right)+\frac{1}{2}\sqrt{\left(\frac{B}{1+A_{1}}+\frac{1}{a^{3(1+A_{1})(1+\alpha)}}\right)^{2}
+\frac{4C(A_{2})}{a^{3(1+A_{1})(1+\alpha)}}},
\end{equation}
this is appropriate solution only at the early universe where $\rho_{c}$ is very big. This is reasonable because terms of $\rho_{c}^{n}$ in ECG are more important at the early universe. If we use relation (\ref{R6}) for the late time, then resulting cosmological parameter are far from observation which will be illustrated graphically in the next sections. So it is desirable to find a solution which yields to good results for both early and late time.\\
There is also alternative way to obtain exact solution. We assume the following conditions,
\begin{eqnarray}\label{ECG3}
\alpha&=&1,\nonumber\\
A_{1}&=&A_{2}-1,\nonumber\\
B&=&2A_{2}.
\end{eqnarray}
Therefore, the only free parameter of the model is $A_{2}$, and we can solve the conservation equation (\ref{HL12}) to obtain the following relation,
\begin{equation}\label{ECG4}
\rho_{c}=c_{2}\frac{(1+x+\sqrt{5x-1})}{(x-1)},
\end{equation}
with $x=c_{1} e^{3\pi}a^{30 A_2}$, where $c_{1}$ is constant of integration and $c_{2}$ comes from the fact that the equation (\ref{ECG2}) is
dimensionless. The density is physically defined only for
$x>1$ (the density is negative for $x<1$). In this model, the universe
starts at $x=1$ with an infinite density, then the density decreases and
finally reaches an asymptotic value for $x\rightarrow +\infty$.\\
In order to write solution (\ref{ECG4}), we used $\tan^{-1}{(\rho_{c}+1)}\approx \pi/2$ approximation, which is valid when $\rho_{c}\gg1$ corresponding to the early universe. However, our solution will be valid at all time and our approximate solution is very close to the late time behavior with $\rho_{c}\ll1$. This is due to the fact that $\tan^{-1}{(\rho_{c}+1)}\approx \pi/4$, for $\rho_{c}\ll1$. Therefore, we will use energy density (\ref{ECG4}) instead of (\ref{R6}) which is only appropriate for the early universe with $\rho_{c}\gg1$. Using numerical analysis will show that energy density (\ref{ECG4}) yields to better results.\\
The equation (\ref{ECG2}) is implicitly normalized by the cosmological
density $\rho_L$, i.e. by the asymptotic value of $\rho_{c}$ for
$a\rightarrow +\infty$. Indeed, we can see that for $\rho_{c}=1$ we get $p_{c}=A_1+A_2-B=-1$ which corresponds to the
equation of state $p=-\rho$ expected for $a\rightarrow +\infty$. According to this
normalization, we must take $c_2=1$. Then, we have to determine $c_1$. The present density and the
cosmological density are related to each other by $\rho_0=1.31\rho_L$.
This means that with the previous normalization we should take
$\rho_{c}=1.31$ when $a=1$. This gives $c_1 e^{(3\pi)}=65$, so we can write $x=65 a^{(30 A_2)}$.\\
Similarly to the detailed balance case, in the IR with $\lambda=1$, $G_{cosmo}=G_{grav}\equiv G$. So, using (\ref{HL8}), (\ref{HL9}) and (\ref{HL10}) one can rewrite Friedmann equations (\ref{HL6}) and (\ref{HL7}) as follows,
\begin{equation}\label{ECG5}
H^{2}=\frac{8\pi G}{3}\rho+\frac{\Lambda}{2}-\frac{k}{a^{2}}+\frac{k^{2}}{2\Lambda a^{4}},
\end{equation}
and,
\begin{equation}\label{ECG6}
\dot{H}+\frac{3}{2}H^{2}=-4\pi G p+\frac{3\Lambda}{4}-\frac{k}{2a^{2}}-\frac{k^{2}}{4\Lambda a^{4}},
\end{equation}
where,
\begin{equation}\label{ECG6-1}
\rho=\rho_{r}+\rho_{b}+\rho_{c},
\end{equation}
is total energy density including radiation, baryonic matter, and extended Chaplygin gas respectively. Therefore,
\begin{equation}\label{ECG7}
p=\frac{1}{3}\rho_{r}+(A_{2}-1)\rho_{c}+A_{2}\rho_{c}^{2}-\frac{2A_{2}}{\rho_{c}},
\end{equation}
is the total pressure with the fact that $\omega_{b}=p_{b}/\rho_{b}=0$, $\omega_{r}=p_{r}/\rho_{r}=1/3$ and $\omega_{c}=p_{c}/\rho_{c}$. Finally it is useful to define the following relations,
\begin{equation}\label{ECG8}
\Omega_{i}=\frac{8\pi G}{3 H^{2}}\rho_{i}, \hspace{5mm} \Omega_{k}=-\frac{k}{a^{2} H^{2}}, \hspace{5mm} \Omega_{0}=\frac{\Lambda}{2 H_{0}^{2}},
\end{equation}
with $i=b, r, c$.
\section{Observational constraints}
\begin{figure}[h!]
 \begin{center}$
 \begin{array}{cccc}
\includegraphics[width=50 mm]{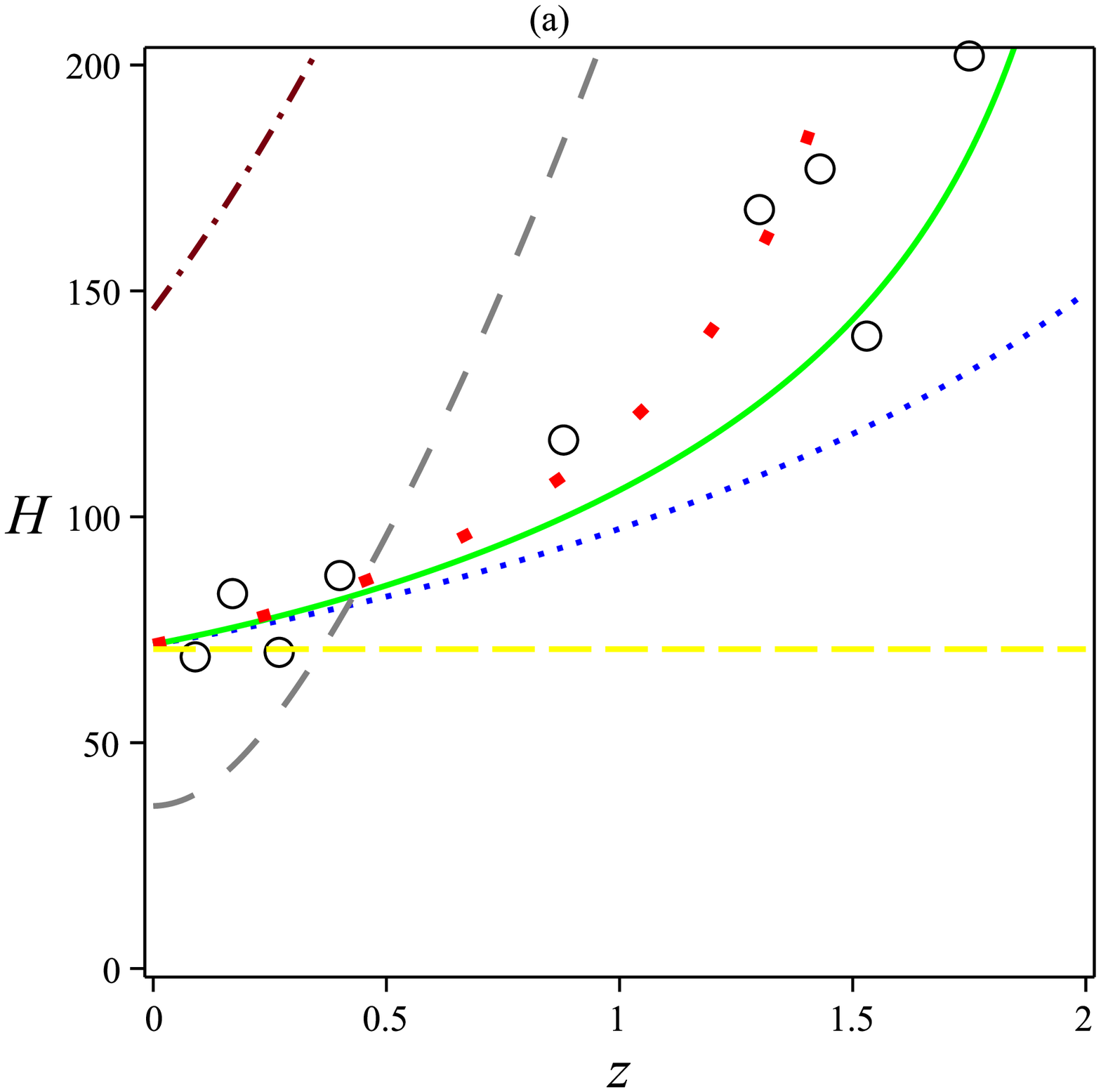}&\includegraphics[width=50 mm]{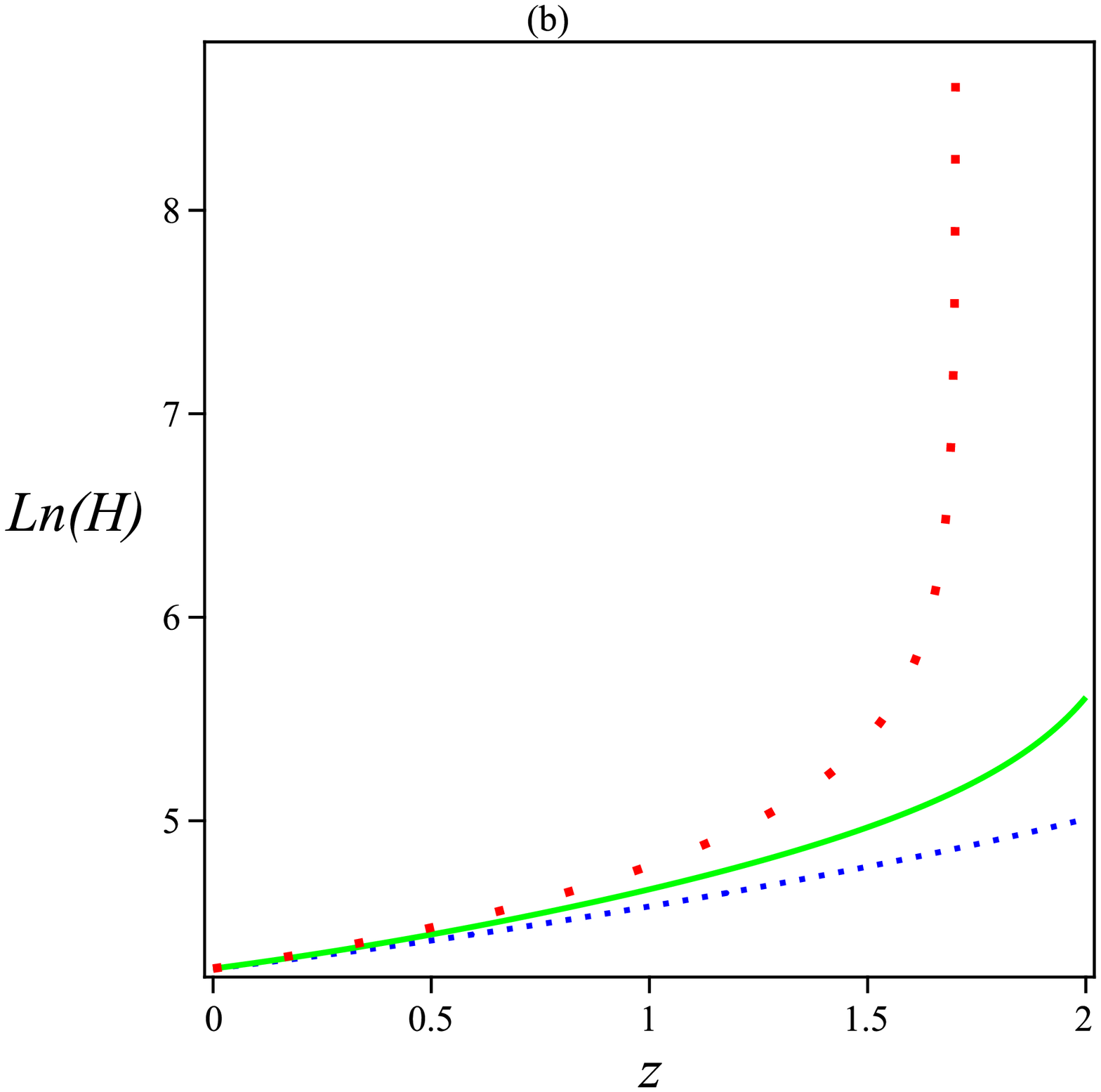}&\includegraphics[width=50 mm]{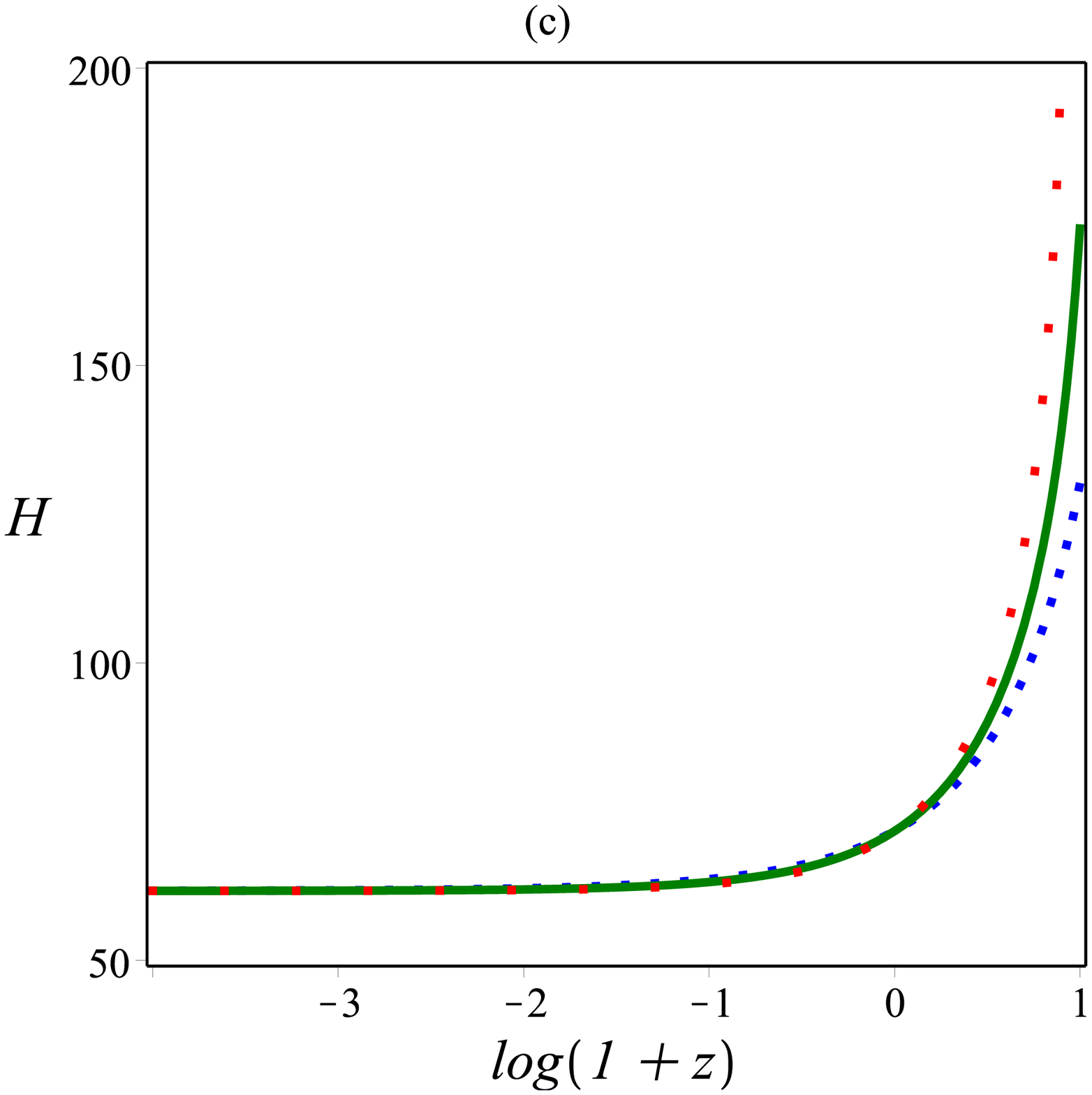}
 \end{array}$
 \end{center}
\caption{(a) Hubble expansion parameter in terms of redshift for the exact solution given by (\ref{ECG4}) with $A_{2}=0.1$ (dotted blue), $A_{2}=0.12$ (solid green), $A_{2}=0.14$ (space dotted red). Dashed yellow ($k=0$), space dashed Gray ($k=1$) and dash dotted ($k=-1$) lines are corresponding to approximate solution given by (\ref{R6}) with $B=0.1$, $C(A_{2})=1$ and $A_{1}=-0.9$. Circles represent $H(z)$ data.  (b) Logarithmic Hubble parameter with $A_{2}=0.1$ (dotted blue) $A_{2}=0.12$ (solid green) $A_{2}=0.14$ (space dotted red). (c) Log scale of Hubble parameter for $1+z$ with $A_{2}=0.1$ (dotted blue), $A_{2}=0.12$ (solid green), $A_{2}=0.14$ (space dotted red).}
 \label{fig1}
\end{figure}

In this section, we use $H(z)$ data to fix some model parameters.
In order to use observational data it is useful to rewrite equations in terms of redshift. In that case, using the following relation,
\begin{equation}\label{O1}
a=\frac{a_{0}}{1+z},
\end{equation}
together with (\ref{HL12}) and (\ref{ECG4}) in the equation (\ref{ECG5}) we can obtain Hubble expansion parameter in terms of redshift as follow,
\begin{equation}\label{O2}
E^{2}(z)=\Omega_{r0}(1+z)^{4}+\Omega_{b0}(1+z)^{3}+\Omega_{c0}F(z)+\Omega_{0}+\Omega_{k0}(1+z)^{2}+\frac{\Omega_{k0}^{2}}{4\Omega_{0}}(1+z)^{4},
\end{equation}
where,
\begin{equation}\label{O3}
E(z)\equiv\frac{H(z)}{H_{0}},
\end{equation}
and,
\begin{equation}\label{O4}
F(z)=\frac{(1+65(1+z)^{-30 A_2}+\sqrt{325(1+z)^{-30 A_2}-1})}{(65(1+z)^{-30 A_2}-1)},
\end{equation}
with the current value of the Hubble expansion parameter $H_{0}$. An important free parameter is $A_{2}$ which should fixed using observational data. Moreover, present day ($z=0$) radiation,
baryon, ECG, cosmological constant, and curvature energy densities are denoted by $\Omega_{r0}$, $\Omega_{b0}$, $\Omega_{c0}$, $\Omega_{0}$, $\Omega_{k0}$ respectively which satisfy the following equation,
\begin{equation}\label{O5}
1=\Omega_{r0}+\Omega_{b0}+\Omega_{c0}+\Omega_{0}+\Omega_{k0}+\frac{\Omega_{k0}^{2}}{4\Omega_{0}}.
\end{equation}
It is obvious that the value of $A_{2}$ is not important at present stage and, as expected, it is important at the early universe.
The last term in the equation (\ref{O5}) corresponds to the dark radiation, which is a characteristic
feature of the HL theory of gravity and restricted as follow \cite{HL-restriction},
\begin{equation}\label{O6}
\frac{\Omega_{k0}^{2}}{4\Omega_{0}}=0.135\Delta N_{\nu}\Omega_{r0},
\end{equation}
where $\Delta N_{\nu}$ represents the effective neutrino species with the following bound \cite{HL-restriction},
\begin{equation}\label{O7}
-1.7\leq\Delta N_{\nu}\leq2.
\end{equation}
However, we restrict ourself to the case of $0\leq\Delta N_{\nu}\leq2$ \cite{HLMCG1}. Using the equation (\ref{O6}) in the relation (\ref{O5}) it is easy to find \cite{HLMCG1},
\begin{equation}\label{O8}
\Omega_{0}=1-\Omega_{b0}-\Omega_{c0}-(1-0.135\Delta N_{\nu})\Omega_{r0}-0.73\sqrt{\Delta N_{\nu}(\Omega_{r0}-(\Omega_{b0}+\Omega_{c0})\Omega_{r0}-\Omega_{r0}^{2})}.
\end{equation}
Then, equation (\ref{O6}) gives,
\begin{equation}\label{O9}
\Omega_{k0}=\sqrt{0.54\Omega_{0}\Delta N_{\nu}\Omega_{r0}}.
\end{equation}
So, in numerical analysis, similar to the Ref. \cite{HLMCG1}, we choose $H_{0}=71.4 \hspace{1mm} Km/s/Mpc$, $\Omega_{b0}=0.04$, $\Omega_{r0}=8.14\times10^{-5}$, and $\Omega_{c0}=0.951$. These yield to $0.0080\leq\Omega_{0}\leq0.0089$ and $0\leq\Omega_{k0}\leq0.00084$. In the Fig. \ref{fig1} we represent our numerical results together experimental data \cite{MNRASoc397(2009)1935}. In order to have more agreement with seven year WMAP data \cite{7Y} we suggests $0.1\leq A_{2}<0.15$ for all $0\leq\Delta N_{\nu}\leq2$. Also, dashed and dotted lines of $H(z)$ show that approximate solution given by (\ref{ECG4}) may lead to unexpected results.

\section{Cosmological parameters}
\begin{figure}[h!]
 \begin{center}$
 \begin{array}{cccc}
\includegraphics[width=60 mm]{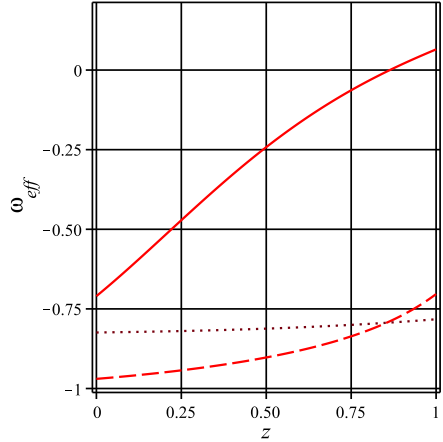}
 \end{array}$
 \end{center}
\caption{EoS in terms of redshift for the exact solution given by (\ref{ECG4}) with $A_{2}=0.15$, open and closed universe denoted by solid line and flat universe denoted by dashed line. Dotted line represent approximate solution given by (\ref{R6}) with $B=0.1$, $C(A_{2})=1$ and $A_{1}=-0.9$ for open and closed universe.}
 \label{fig5}
\end{figure}

Using the obtained results for model parameters we investigate behavior of some important cosmological parameters. The first quantity is the effective EoS parameter given by,
\begin{equation}\label{Co1}
\omega_{eff}=\frac{p_{eff}}{\rho_{eff}},
\end{equation}
where,
\begin{equation}\label{Co2}
p_{eff}=p+\frac{2}{\kappa^{2}}\left[\frac{k^{2}}{\Lambda a^{4}}-3\Lambda\right],
\end{equation}
and,
\begin{equation}\label{Co3}
\rho_{eff}=\rho+\frac{2}{\kappa^{2}}\left[\frac{3k^{2}}{\Lambda a^{4}}+3\Lambda\right],
\end{equation}
with the $p$ and $\rho$ given by the equations (\ref{ECG6-1}) and (\ref{ECG7}). In the Fig. \ref{fig5} we can see behavior of the EoS parameter for open, close and flat universe. In the case of flat universe we see $\omega_{eff}\rightarrow-1$ while in the cases of closed or open universes, the
present value of the effective EoS parameter is about -0.7 which admits accelerating universe in agreement with observational data \cite{RM2002}.\\
In order to have a fair evolution of the model it is desirable to compare the equation of state parameter with the current bounds on $\omega(a)=\frac{\omega_{eff}}{\Omega_{c}}$.
As we know, there are several ways to parameterize the equation of state parameter for example Chevallier-Polarski-Linder (CPL) parameterizations \cite{CP,L},
\begin{equation}\label{CPL}
\omega(a)=\omega_{0}+\omega_{a}(1-a)=\omega_{0}+\omega_{a}\frac{z}{1+z},
\end{equation}
where $\omega_{0}$ corresponds to the present day value and \cite{Olga}
\begin{equation}\label{wa}
\omega_{a} = (d\omega(a)/dz)|z=0 =
(-d\omega(a)/da)|a=1.
\end{equation}
The current constraints for the CPL parameters are $\omega_{0} =  -1.11 \pm 0.17$ and $\omega_{a}= 0.34 \pm 0.60$ \cite{last}. We can summarize our results using present value $\Omega_{c0}=0.951$ \cite{HLMCG1} in the table 1 and see agreement with current data.

\begin{center}
  \begin{tabular}{|@{} l @{} ||@{} c @{} || @{}r @{}|}
    \hline
    $A_{2}$ & $\omega_{0}$ & $\omega_{a}$ \\ \hline\hline
    0.11    & -1.02660     & 0.05699       \\ \hline
    0.13    & -1.02300     & 0.07297       \\ \hline
    0.15    & -1.01960     & 0.09158       \\ \hline
    0.17    & -1.01619     & 0.11282       \\ \hline
    0.19    & -1.01270     & 0.13680       \\
    \hline
  \end{tabular}
\end{center}
{\it{Table 1. Values of $\omega_{0}$ and $\omega_{a}$  of our model with $\Lambda=\kappa=1, k=0$ and different values of $A_{2}$, in agreement with $\omega_{0} =  -1.11 \pm 0.17$ and $\omega_{a}= 0.34 \pm 0.60$ \cite{last}.}}\\\\\\\\\\\\

\begin{figure}[h!]
 \begin{center}$
 \begin{array}{cccc}
\includegraphics[width=60 mm]{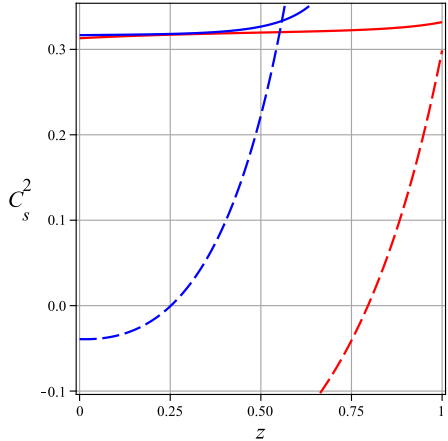}
 \end{array}$
 \end{center}
\caption{Squared sound speed in terms of redshift for $A_{2}=0.15$ (red line), $A_{2}=0.2$ (blue line). Open and closed universe denoted by solid line and flat universe denoted by dashed line.}
 \label{fig6}
\end{figure}

We can also investigate stability of the model using sound speed which is given by,
\begin{equation}\label{Co4}
C_{s}^{2}=\frac{dp}{d\rho},
\end{equation}
in which, the model is stable if $C_{s}^{2}\geq0$. In the Fig. \ref{fig6} we can see that the model is completely stable for open and closed universes, while there are some instabilities in the flat universe for $0\leq A_{2}\leq0.2$.

\section{Thermodynamics}
It is important to investigate thermodynamic properties of the model which may lead to study the generalized second law of thermodynamic.
In order to do that we consider the universe as a thermodynamical system with the apparent horizon surface being its
boundary. The apparent horizon given by \cite{TBH5},
\begin{equation}\label{T01}
r_{A}=\left(H^{2}+\frac{k}{a^{2}}\right)^{-\frac{1}{2}}.
\end{equation}
Then, the temperature and the entropy are given by the following expressions respectively,
\begin{equation}\label{T02}
T=\frac{1}{2\pi r_{A}},
\end{equation}
and,
\begin{equation}\label{T03}
S=\frac{\kappa^{2}}{32\Lambda G^{2}}\left[\Lambda r_{A}^{2}+2k \ln{(\sqrt{\Lambda}r_{A})}\right].
\end{equation}
We should have mentioned that the first term corresponds
to the general relativity, while the second one arises from HL gravity.
Using the equation (\ref{ECG5}) we can obtain,
\begin{equation}\label{T04}
T=\frac{\sqrt{\frac{8\pi G}{3}\rho+\frac{\Lambda}{2}+\frac{k^{2}}{2\Lambda a^{4}}}}{2\pi},
\end{equation}
and,
\begin{equation}\label{T05}
S=\frac{\kappa^{2}}{32\Lambda G^{2}}\left[\frac{\Lambda}{\frac{8\pi G}{3}\rho+\frac{\Lambda}{2}+\frac{k^{2}}{2\Lambda a^{4}}} +2k \ln{(\sqrt{\frac{\Lambda}{\frac{8\pi G}{3}\rho+\frac{\Lambda}{2}+\frac{k^{2}}{2\Lambda a^{4}}}})}\right].
\end{equation}
As we can see from the equation (\ref{T03}), the apparent horizon is a function of Hubble parameter, so it is a function of time. Therefore, $dr_{A}$ is corresponding to $dV$. In that case, the first law of thermodynamics \cite{TBH5} given by,
\begin{equation}\label{T06}
TdS=dE+pdV,
\end{equation}
where $V = \frac{4}{3}\pi r_{A}^{3}$ is the volume of the system bounded
by the apparent horizon, so $dV$ denotes volume-change and $dE=\rho dV$, where $E=\frac{4}{3}\pi r_{A}^{3}\rho$ is energy. So, one can use (\ref{HL11}), (\ref{HL11-1}), and (\ref{HL12}) in the equation (\ref{T06}), and obtain,
\begin{equation}\label{T08}
\frac{dS_{2}}{dt}=\frac{4\pi}{T}(1+\omega)\rho r_{A}^{2}(\frac{dr_{A}}{dt}-Hr_{A}).
\end{equation}
where $S_{2}$ is the entropy given by the first law of thermodynamics (\ref{T06}).
Using the relation (\ref{T02}) gives,
\begin{equation}\label{T09}
\frac{dS_{2}}{dt}=2(1+\omega)\rho r_{A}(\frac{dr_{A}}{dt}-Hr_{A}).
\end{equation}
In that case one can relate the horizon entropy to $r_{A}$. It has been constructed using the relation (\ref{T03}) \cite{Bh2, TBH2, TBH3}.
Differentiating (\ref{T03}) we have,
\begin{equation}\label{T10}
\frac{dS_{1}}{dt}=\frac{\kappa^{2}}{16\Lambda G^{2}}\left[\Lambda r_{A} +\frac{k}{r_{A}} \right]\frac{dr_{A}}{dt},
\end{equation}
where $S_{1}$ is the entropy given by the apparent
horizon.  As expected, the first term is useful while the second term coming from HL gravity. Now, we can calculate total entropy change,
\begin{equation}\label{T11}
\frac{dS_{tot}}{dt}=\frac{dS_{1}}{dt}+\frac{dS_{2}}{dt}.
\end{equation}
It is easy to check that $\frac{dS_{tot}}{dt}>0$, at least for flat and closed universe, in agreement with the results reported by Ref. \cite{TBH5}. In the case of open universe ($k=-1$), it should be $r_{A}^{2}\geq\frac{1}{\Lambda}$, to have $\frac{dS_{1}}{dt}\geq0$.\\
We have the following conditions to have conserved entropy ($\frac{dS_{tot}}{dt}=0$),
\begin{eqnarray}\label{CE}
\frac{dr_{A}}{dt}&=&Hr_{A}, \nonumber\\
r_{A}^{2}&=&-\frac{k}{\Lambda}.
\end{eqnarray}
Both conditions are satisfied simultaneously, only in the case of varying $\Lambda$ (slowly) as follow,
\begin{equation}\label{Lambda}
\Lambda=\pm Ce^{-2\int{Hdt}},
\end{equation}
for the closed and open universe. Otherwise, in the case of constant $\Lambda$, total entropy is not conserved.

\section{Conclusion}
In this paper, we have considered the extended Chaplygin gas in Horava-Lifshitz gravity. First of all we reviewed the HL cosmology and then solved conservation equation for the special case of the ECG at the second order. Also, we solved the general case approximately and obtained energy density in terms of scale factor which allow us to investigate Hubble expansion parameter in terms of redshift. We have considered some assumptions which reduced free parameters of the ECG to one. We used $H(z)$ data to constrain this parameter in HL
cosmology. $H(z)$ observations suggest that $0.13\leq A_{2}\leq0.25$, which means $0.26\leq B\leq0.5$ and $-0.87\leq A_{1}\leq-0.75$. Using the obtained parameters of the model we have studied evolution of EoS parameter and found that, for the both cases $k<0$ and $k>0$, the present value of the effective EoS parameter is about -0.7, while it yields to 1/3 at high redshift. On the other hand, for the flat universe ($k=0$) we found $\omega_{eff}\rightarrow-1$.\\
Stability of the model is also investigated by using the squared sound speed and found that the model is completely stable for the open and closed universes, while there are some instabilities in flat universe for the case of $0.1\leq A_{2}\leq0.2$. It seems $A_{2}=0.15$ is the best fitted value.\\
Finally, we have investigated thermodynamic properties of the model. We found that the generalized second law of thermodynamics is valid for a flat or closed universe. We found that the total entropy may be conserved in the case of varying $\Lambda$.\\
In this work we have neglected the quantum effects near the black hole horizon, so it would be interesting to consider such effects for example using logarithmic corrected entropy \cite{Log1}. Finally an interesting work may be the consideration of unified Chaplygin gas with the following equation of state,
\begin{equation}\label{unified}
p=\sum_{i\in R}A_{i}\rho^{i},
\end{equation}
where $i$ may be positive, negative, integer and non-integer number. It covers all kinds of Chaplygin gas EoS and will be investigate as separate work.
\\\\

\begin{acknowledgments}
Author would like to thank P. H. Chavanis and E.O. Kahya for useful comments and discussions. Also special thanks to Hoda Farahani for reading manuscript and helpful discussions.
\end{acknowledgments}

\end{document}